\documentclass[11pt]{article}
\usepackage{times}
\usepackage{sidecap}
\usepackage{geometry}
\usepackage{floatrow}
\usepackage{float}
\usepackage{hyperref}
\usepackage{graphics,graphicx}
\usepackage{subcaption}
\usepackage{array}
\usepackage{epsf,wrapfig}

\usepackage{natbib,bm}

\usepackage{paralist}
\let\olditem\item
\renewenvironment{thebibliography}[1]{%
  \section*{\refname}
  \let\par\relax
  \renewcommand{\item}[1][]{\olditem[\textbullet]}%
  \inparaenum}{\endinparaenum}

\newcounter{aff}

\newcommand{\statS}[2]{\normalsize WSU (#1$\times$) $\rightarrow$ 2040 (#2$\times$)}

\newcommand{\statBW}[2]{\normalsize WSU (#1$\times$) $\rightarrow$ 2040 (#2$\times$)}

\newcommand{\statQ}[2]{\normalsize WSU (#1) $\rightarrow$ 2040 (#2)}

\DeclareGraphicsExtensions{.jpg,.pdf,.png,.eps}

\usepackage{caption}
\usepackage{xcolor}

\bibpunct{(}{)}{;}{a}{}{,}

\usepackage{soul}

\definecolor{darkcyan}{rgb}{0.0, 0.0, 0.7}

\definecolor{boxcolor}{RGB}{193, 229, 255}
\definecolor{figboxcolor}{RGB}{193, 229, 255}
\definecolor{boxcolorgray}{RGB}{227, 227, 227}
\definecolor{darkcyan}{rgb}{0.0, 0.0, 0.7}

\sethlcolor{boxcolorgray}

\def\gtrsim{\lower 2pt \hbox{$\, \buildrel {\scriptstyle >}\over
{\scriptstyle \sim}\,$}}
\def\lesssim{\lower 2pt \hbox{$\, \buildrel {\scriptstyle <}\over
{\scriptstyle \sim}\,$}}

\def\approxlt{\lower.2em\hbox{$\buildrel < \over \sim$}}
\def\approxgt{\lower.2em\hbox{$\buildrel > \over \sim$}}

\def\be{\begin{equation}}
\def\ee{\end{equation}}
\def\bea{\begin{eqnarray*}}
\def\eea{\end{eqnarray*}}

\def\jref@jnl#1{{\rm#1}}
\def\aj{\jref@jnl{AJ}}			
\def\araa{\jref@jnl{ARA\&A}}		
\def\apj{\jref@jnl{ApJ}}			
\def\apjl{\jref@jnl{ApJ}}		
\def\apjs{\jref@jnl{ApJS}}		
\def\ao{\jref@jnl{Appl.~Opt.}}		
\def\apss{\jref@jnl{Ap\&SS}}		
\def\aap{\jref@jnl{A\&A}}		
\def\aapr{\jref@jnl{A\&A~Rev.}}		
\def\aaps{\jref@jnl{A\&AS}}		
\def\azh{\jref@jnl{AZh}}			
\def\baas{\jref@jnl{BAAS}}		
\def\jrasc{\jref@jnl{JRASC}}		
\def\memras{\jref@jnl{MmRAS}}		
\def\mnras{\jref@jnl{MNRAS}}		
\def\pra{\jref@jnl{Phys.~Rev.~A}}	
\def\prb{\jref@jnl{Phys.~Rev.~B}}	
\def\prc{\jref@jnl{Phys.~Rev.~C}}	
\def\prd{\jref@jnl{Phys.~Rev.~D}}	
\def\pre{\jref@jnl{Phys.~Rev.~E}}	
\def\prl{\jref@jnl{Phys.~Rev.~Lett.}}	
\def\pasp{\jref@jnl{PASP}}		
\def\pasj{\jref@jnl{PASJ}}		
\def\qjras{\jref@jnl{QJRAS}}		
\def\skytel{\jref@jnl{S\&T}}		
\def\solphys{\jref@jnl{Sol.~Phys.}}	
\def\sovast{\jref@jnl{Soviet~Ast.}}	
\def\ssr{\jref@jnl{Space~Sci.~Rev.}}	
\def\zap{\jref@jnl{ZAp}}			
\def\nat{\jref@jnl{Nature}}		
\def\iaucirc{\jref@jnl{IAU~Circ.}}
\def\aplett{\jref@jnl{Astrophys.~Lett.}}
\def\apspr{\jref@jnl{Astrophys.~Space~Phys.~Res.}}
\def\bain{\jref@jnl{Bull.~Astron.~Inst.~Netherlands}}
\def\fcp{\jref@jnl{Fund.~Cosmic~Phys.}}
\def\gca{\jref@jnl{Geochim.~Cosmochim.~Acta}}
\def\grl{\jref@jnl{Geophys.~Res.~Lett.}}
\def\jcp{\jref@jnl{J.~Chem.~Phys.}}	
\def\jgr{\jref@jnl{J.~Geophys.~Res.}}	
\def\jqsrt{\jref@jnl{J.~Quant.~Spec.~Radiat.~Transf.}}
\def\memsai{\jref@jnl{Mem.~Soc.~Astron.~Italiana}}
\def\nphysa{\jref@jnl{Nucl.~Phys.~A}}
\def\physrep{\jref@jnl{Phys.~Rep.}}
\def\physscr{\jref@jnl{Phys.~Scr}}
\def\planss{\jref@jnl{Planet.~Space~Sci.}}
\def\procspie{\jref@jnl{Proc.~SPIE}}

\def\ion#1#2{#1$\;${\small\rm\@Roman{#2}}\relax}

\newcount\lecurrentfam
\def\LaTeX{\lecurrentfam=\the\fam \leavevmode L\raise.42ex
\hbox{$\fam\lecurrentfam\scriptstyle\kern-.3em A$}\kern-.15em\TeX}
\def\sizrpt{
(\fontname\the\font): em=\the\fontdimen6\font, ex=\the\fontdimen5\font
\typeout{
(\fontname\the\font): em=\the\fontdimen6\font, ex=\the\fontdimen5\font
}}
\definecolor{dark}{rgb}{ 0.34, 0.09, 0.49}

\usepackage{fancyhdr,url}
\usepackage{transparent}
\usepackage{tablefootnote}
\usepackage[most]{tcolorbox}
\usepackage{mathpazo}

\begin{document}

% Title and subtitle
\noindent
\begin{raggedright}
{\Huge\bfseries Understanding cosmic explosions \\[0.5em]
\Large How do shocks, jets, and mass-loss processes operate in the most extreme environments, and what physics is accessible only at millimeter wavelengths?
}\end{raggedright}\\

%\vfill
% Author block
%\thispagestyle{empty}
%\vspace{1em}

\noindent Karri I. I. Koljonen$^{1,2}$, 
Tomek Kami\'nski$^{3}$, 
Anna Y. Q. Ho$^{4}$, 
Thomas Maccarone$^{5}$, 
Aneurin Evans$^{6}$,  
Mar\'ia D\'iaz Trigo$^{7}$, 
Thomas D. Russell$^{8}$, 
Dieu D. Nguyen$^{9}$, 
Dirk Petry$^{7}$,
Jakob van den Eijnden$^{10}$,
Tanmoy Laskar$^{11}$,
Laura Chomiuk$^{12}$,
and the ALMA2040 Transient and Time-Domain Astronomy working group

\vspace{1em}
{
\footnotesize{
\noindent 
$^{1}$ Institut for Fysikk, Norwegian University of Science and Technology, H{\o}gskloreringen 5, Trondheim, 7491, Norway \\
$^{2}$ Finnish Centre for Astronomy with ESO (FINCA), FI-20014 University of Turku, Finland \\
$^{3}$ Nicolaus Copernicus Astronomical Center, Polish Academy of Sciences, Rabia\'nska 8, 87-100 Toru\'n, Poland \\
$^{4}$ Department of Astronomy, Cornell University, Ithaca, NY 14853, USA \\
$^{5}$ Department of Physics \& Astronomy, Texas Tech University, Box 41051, Lubbock, TX, 79409-1051, USA \\
$^{6}$ Astrophysics Group, Lennard Jones Laboratory, Keele University, Keele, Staffordshire ST5 5BG, UK \\
$^{7}$ European Southern Observatory (ESO), Karl-Schwarzschild-Strasse 2, D-85748 Garching bei München, Germany \\
$^{8}$ INAF, Istituto di Astrofisica Spaziale e Fisica Cosmica, via U. La Malfa 153, I-90146 Palermo, Italy \\
$^{9}$ Department of Astronomy, University of Michigan, 1085 South University Avenue, Ann Arbor, MI 48109, USA\\
$^{10}$ Anton Pannekoek Institute for Astronomy, Universiteit van Amsterdam, Science Park 904, NL-1098 XH, Amsterdam, the Netherlands \\
$^{11}$ Department of Physics \& Astronomy, University of Utah, Salt Lake City, UT 84112, USA\\
$^{12}$ Center for Data Intensive and Time Domain Astronomy, Department of Physics and Astronomy, Michigan State University, East Lansing, MI 48824, USA \\
}
}
\begin{tcolorbox}[boxsep=3pt,left=0pt,right=0pt,top=1pt,bottom=1pt,colframe=boxcolor,colback=boxcolor]
%
%
%
%\vfill
%\begin{center} \transparent{0.8}
%\includegraphics[width=0.8\textwidth]{51c17472e095aeda42f860d5241b9e8eb384e7e7.jpg} \\% Opacity
%{\it Image credit: Jeff Cooke/Swinburne University of Technology}
%\end{center}
%%\vspace{2em}
%
%\vfill
%{
%\begin{flushleft}
%\textbf{Keywords:} transients, Techniques: interferometric, Stars: binary, Stars: supernovae, Stars: novae
%\end{flushleft}
%}
\justify
\textbf{Abstract:} Millimeter-wavelength observations provide a uniquely sensitive window into the physics of cosmic explosions, from supernovae (SNe) and tidal disruption events (TDEs) to nova eruptions, stellar mergers and electromagnetic counterparts to gravitational wave (GW) sources. Over the past decade, discoveries across the transient sky have revealed that many of the most informative phases of shock interaction, particle acceleration and jet formation unfold at frequencies above the traditional cm-wave domain, where synchrotron self-absorption weakens and thermal, nonthermal and dust components can be disentangled. Upcoming wide-field mm surveys will detect a new population of short-lived transients but will be limited by coarse localization and limited spectral characterization. ALMA2040, with dramatically improved sensitivity, wide instantaneous spectral coverage, rapid-response and sub-arraying modes, polarimetry and extended baselines, is essential for converting these discoveries into robust physical constraints. Here we outline the scientific context, key open questions and the ALMA2040 capabilities required to unlock the full potential of millimeter time-domain astrophysics in the 2040s.
\end{tcolorbox}

\newpage{}

\section{Scientific Context and Open Questions}

%\newpage
%% \vspace{-18pt}
%\begin{figure} %\vspace{-1cm}
%\floatbox[{\capbeside\thisfloatsetup{capbesideposition={right,top},capbesidewidth=9.1cm}}]{figure}[\FBwidth]
%{\caption{\it Placeholder text and image, verbosely expanded to span the full width of the page. 
%\vspace{-0.7cm}\vspace{-0.5cm}}
%}
%{\includegraphics[width=7cm]{composite_spectrum.png}}
%\vspace{-0.5cm}
%\end{figure}

Cosmic explosions are diverse laboratories for extreme physics, producing shocks, relativistic outflows, rapid mass loss, nucleosynthesis, and in some cases compact-object formation in real time. Historically, much of what we know about the interaction between explosive ejecta and the surrounding medium has come from modeling centimeter-wavelength synchrotron emission \citep{alexander20,chomiuk21,weiler02,chandra12}. Over the past decade, however, observational breakthroughs have demonstrated that the earliest and most diagnostically powerful phases of these interactions frequently occur at millimeter and sub-millimeter wavelengths, where synchrotron self-absorption can be weaker, free--free opacity can drop rapidly with frequency, and thermal, non-thermal and dust components can be separated with broad spectral coverage \citep[e.g.,][]{ho19,laskar19}.

\begin{tcolorbox}[title=Why ALMA2040 and why mm/sub-mm?, colback=blue!5, colframe=blue!60!black]
By the 2040s, facilities such as SKAO and ngVLA will provide transformational sensitivity at cm wavelengths, enabling late-time calorimetry once synchrotron emission is optically thin and tracing absorption at low frequencies. However, many key diagnostics of cosmic explosions occur at \textit{higher} frequencies and at \textit{earlier} times, when cm emission can be self-absorbed or suppressed by free--free opacity, and when thermal, non-thermal and dust components overlap. Across the transient zoo, ALMA2040 will enable a uniform, quantitative answer to a small set of recurring physical questions that are difficult to address with current mm facilities or with cm-band data alone:

\vspace{0.1cm} \noindent \textbf{Environment:} What are the density scale, radial profile and degree of clumping/asymmetry of the circumstellar/circumnuclear medium (CSM) on radii $\sim10^{14}$--$10^{17}$\,cm, as inferred from the early-time spectral peak and its evolution?

\vspace{0.1cm} \noindent \textbf{Energetics:} What are the shock/outflow kinetic energies and velocity distributions (including mildly relativistic components) at the earliest phases, before cm emission becomes optically thin?

\vspace{0.1cm} \noindent \textbf{Absorption physics:} Is the early turnover set by synchrotron self-absorption, free--free absorption, or a multi-component spectrum, and how does this partition vary across source classes (dense-CSM SNe, novae, FBOTs, TDEs, GRBs)?

\vspace{0.1cm} \noindent \textbf{Shock microphysics:} Do collisionless shocks in transients obey a universal electron distribution and microphysical parameter set, or is there measurable spectral curvature/break evolution indicating time-dependent acceleration/cooling or multiple particle populations?

\vspace{0.1cm} \noindent \textbf{Geometry:} What is the angular structure of jets/outflows (quasi-spherical vs.\ stratified vs.\ collimated), and how do viewing-angle effects map onto observed diversity, as constrained by polarization, spectral evolution and (for the nearest events) size/expansion measurements?

\vspace{0.1cm} \noindent \textbf{Dust/molecules and obscuration:} When and where do dust and molecules form in explosions and mergers, and how do they coexist with shock-powered non-thermal emission in heavily obscured systems?

\vspace{0.1cm} \noindent \textbf{Multi-messenger inference:} For neutrino- and GW-triggered transients, what are the shock energetics, densities and magnetic fields that set the efficiency of hadronic interactions and/or late-time energy injection, and how do these measurements reduce degeneracies in neutrino/GW counterpart models?
\end{tcolorbox}

\begin{figure}
    \centering
    \includegraphics[width=1.0\linewidth]{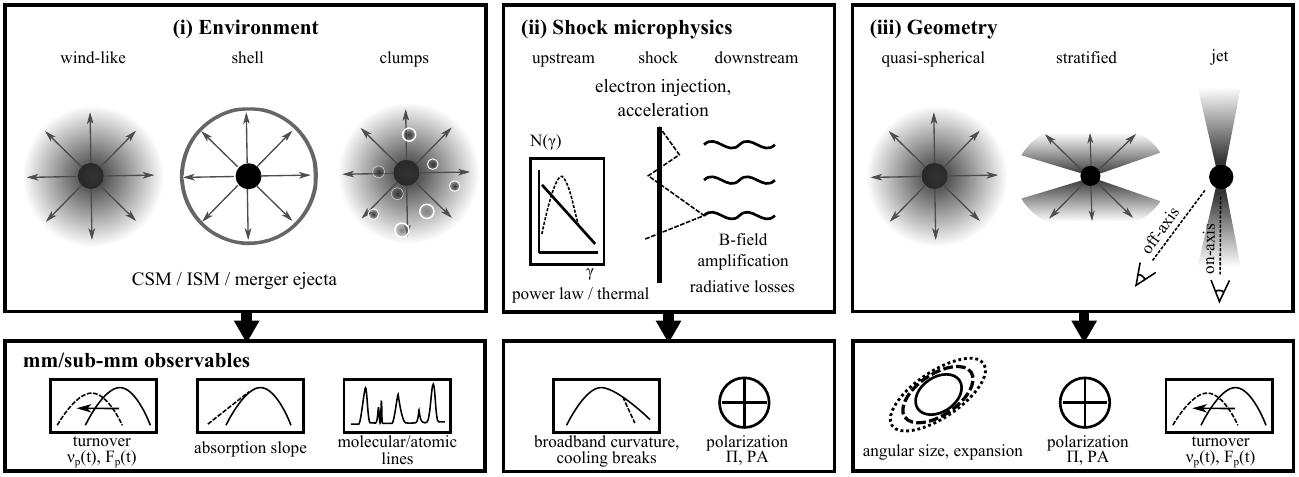}
    \caption{Schematic illustrating three recurring physical effects that shape mm/sub-mm emission from cosmic explosions and the corresponding observables. \textit{Panel (i):} Density structure and composition of the surrounding medium (e.g., wind-like profiles, shells, clumps or merger ejecta) regulate absorption and the evolution of the spectral peak. \textit{Panel (ii)} Electron injection/acceleration, magnetic-field amplification and cooling can imprint broadband spectral curvature and influence polarization properties. \textit{Panel (iii):} Outflow geometries (quasi-spherical, stratified or collimated jets viewed on/off-axis) can be constrained through polarization and direct size/expansion measurements from high-resolution interferometric imaging.}
    \label{fig:effects}
\end{figure}

\noindent Although cosmic explosions span diverse progenitors and engines, their mm emission often reflects the interplay of three recurring physical effects (Figure \ref{fig:effects}): \textit{i)} the density structure and composition of the surrounding material, \textit{ii)} the microphysics of collisionless shocks, and \textit{iii)} the geometry of the outflow (quasi-spherical, stratified, or collimated into a jet, viewed at an arbitrary angle). The mm/sub-mm band provides direct access to observables that map onto these effects at early times, when diagnostic information is otherwise hidden by absorption at longer wavelengths or by reprocessing at shorter wavelengths \citep[e.g.,][]{ho19,laskar19,eftekhari22}. The evolution of the spectral turnover and peak flux constrains both the density scale and radial profile of the ambient medium and the shock energetics  \citep[e.g.,][]{chevalier98,weiler02,Zauderer2011,berger12,ho19,Maeda2021,Nayana2025}. Simultaneous broadband spectral energy distributions (SEDs) across $\sim$35--950~GHz allow tests of whether a single synchrotron component with a standard power-law electron distribution can fit the data, and can reveal time-variable or multi-component behavior \citep[e.g.,][]{ho19,Ho2022}. Polarimetry probes magnetic field ordering and jet magnetization \citep[e.g.,][]{Granot2003,Gill2018}, and high-resolution imaging or size constraints (when feasible) provide an independent handle on the physical radius and expansion speed of the emitting region, reducing model degeneracies that cannot be broken with light curves alone \citep[e.g.,][]{Taylor2004,Mooley2018,margalit24}. ALMA2040 directly targets these observables by combining wideband continuum and spectral-line capability with rapid-response operations, polarimetry and long-baseline imaging, enabling a uniform physical characterization of fast and obscured explosions across the full mm/sub-mm window.

\vspace{0.1cm} \noindent \textbf{Supernovae: interaction-powered and engine-driven (magnetar-powered SLSNe)\,\,} A new appreciation for \emph{``interaction-powered''} SNe has emerged from wide-field optical surveys, which indicate that a substantial fraction of massive stars explode inside dense circumstellar environments shaped within only a few hundred AU of the progenitor \citep{Smith2017,Bruch2021}. The physical origin of this mass loss, whether binary interaction \citep{Chevalier2012}, instabilities in late nuclear burning \citep{Quataert2012}, or both \citep{Wu2022}, remains uncertain. High densities that power optical signatures of interaction often render the earliest shocks opaque at cm wavelengths, while mm emission can escape. Only a small number of such SNe have well-sampled mm observations so far, but these already show the promise of recovering shock properties and the radial structure of the CSM during the first days to weeks \citep{Maeda2021,Nayana2025}. Key open questions include whether dense CSM is smooth or clumpy, whether it is axisymmetric (as expected in binaries), and what fraction of interaction-powered explosions launch mildly relativistic components that would be missed or mischaracterized at cm wavelengths. A complementary engine-driven channel may be provided by magnetar-powered superluminous SNe (SLSNe), which can produce luminous non-thermal radio/mm emission on months--years timescales; mm observations then probe energy injection, nebular magnetization and the surrounding density structure \citep[e.g.,][]{Mukhopadhyay2026}, a connection made timely by recent GeV detections of an SLSN with \emph{Fermi}-LAT \citep{Acero2026}.

\vspace{0.1cm} \noindent \textbf{Engine-powered explosions (TDEs, GRBs, FBOTs)\,\,} Engine-powered explosions show strong and often unexpected behavior in the mm/sub-mm regime. Relativistic jetted TDEs \citep[][]{berger12,rhodes25}, rapidly evolving fast blue optical transients \citep[FBOTs; such as AT2018cow][]{ho19}, and late-time radio/mm rebrightenings in optically selected TDEs \citep{cendes24} can peak at mm frequencies on week-to-month timescales. The underlying mechanisms (reverse shocks, stratified ejecta, magnetic dissipation or multiple electron populations) remain poorly constrained. Existing mm observations already show variable sub-mm spectral indices \citep{ho19}, steep mm spectra inconsistent with standard single power-law electron distribution \citep{Ho2022}, and evidence for multi-component outflow structures. These features motivate simultaneous, wideband mm monitoring to distinguish intrinsic spectral curvature from variability and absorption and to support physical outflow models \citep[e.g.,][]{Mukhopadhyay2024}.

\vspace{0.1cm} \noindent \textbf{Accreting compact objects (XRBs)\,\,}
Accreting stellar-mass compact objects provide a complementary, nearby laboratory for the same physics. In X-ray binaries (XRBs), compact jets in the hard state produce flat-to-inverted synchrotron spectra that can extend into the mm/sub-mm band, while outbursts launch discrete ejecta whose spectral evolution and polarization trace particle acceleration, magnetic field ordering and jet composition on minute--day timescales \citep[e.g.,][]{Fender2004,Gallo2003,Tetarenko2017,Koljonen2018}. XRB outbursts offer a fast, repeatable analogue of engine-powered transient outflows, enabling controlled tests of how jet launching and shock microphysics depend on accretion state and environment (see also the companion ALMA2040 white paper \textit{mm-jets in the multi-messenger era}). Extending such studies to the Magellanic Clouds and, for the brightest events, to the Local Volume will require the sensitivity, rapid response and multi-band capability of ALMA2040, and will provide a unique bridge between Galactic jet phenomenology and the extragalactic explosions discussed above.

\vspace{0.1cm} \noindent \textbf{Classical novae\,\,}
Classical novae remain challenging at high frequencies because their radio/mm emission often combines thermal free-free emission, shock-powered synchrotron and dust components whose relative contributions can evolve rapidly \citep{Nirupam,Diaz}. Shocks strongly influence both the eruptive evolution and the structure of the remnant \citep[e.g.,][]{Deredzinski,Chomiuk}, but disentangling shock-powered synchrotron emission from thermal ejecta requires broad spectral coverage at early times, when the emission transitions from optically thick to thin across the mm band. Robust separation of these components is essential for accurate ejecta-mass and kinetic-energy estimates and for quantifying nova contributions to isotope synthesis on Galactic scales (e.g., $^{7}$Li, $^{13}$C, and $^{22}$Na).

\vspace{0.1cm} \noindent \textbf{Stellar mergers (LRNe and SPRITES)\,\,}
Luminous red novae (LRNe) and the obscured SPRITES population has emerged as a key probe of binary evolution and dust production \citep[e.g.,][]{SPRITES,Kaminski2026}. Millimeter interferometry has revealed complex remnant structures (tori, disks, equatorial winds, jet-like outflows), indicating strong departures from spherical mass loss and directly constraining angular-momentum redistribution during coalescence \citep{KamiPN}. Sensitivity limits currently confine detailed studies to a handful of nearby events; larger samples are needed to map merger diversity and dust/molecule formation channels. Many of these aspects can be investigated only at mm wavelengths, as the dust-enshrouded transients often do not have optical counterparts \citep{SPRITES}. 

\begin{figure}[h!]
    \centering
    \includegraphics[width=1.0\linewidth]{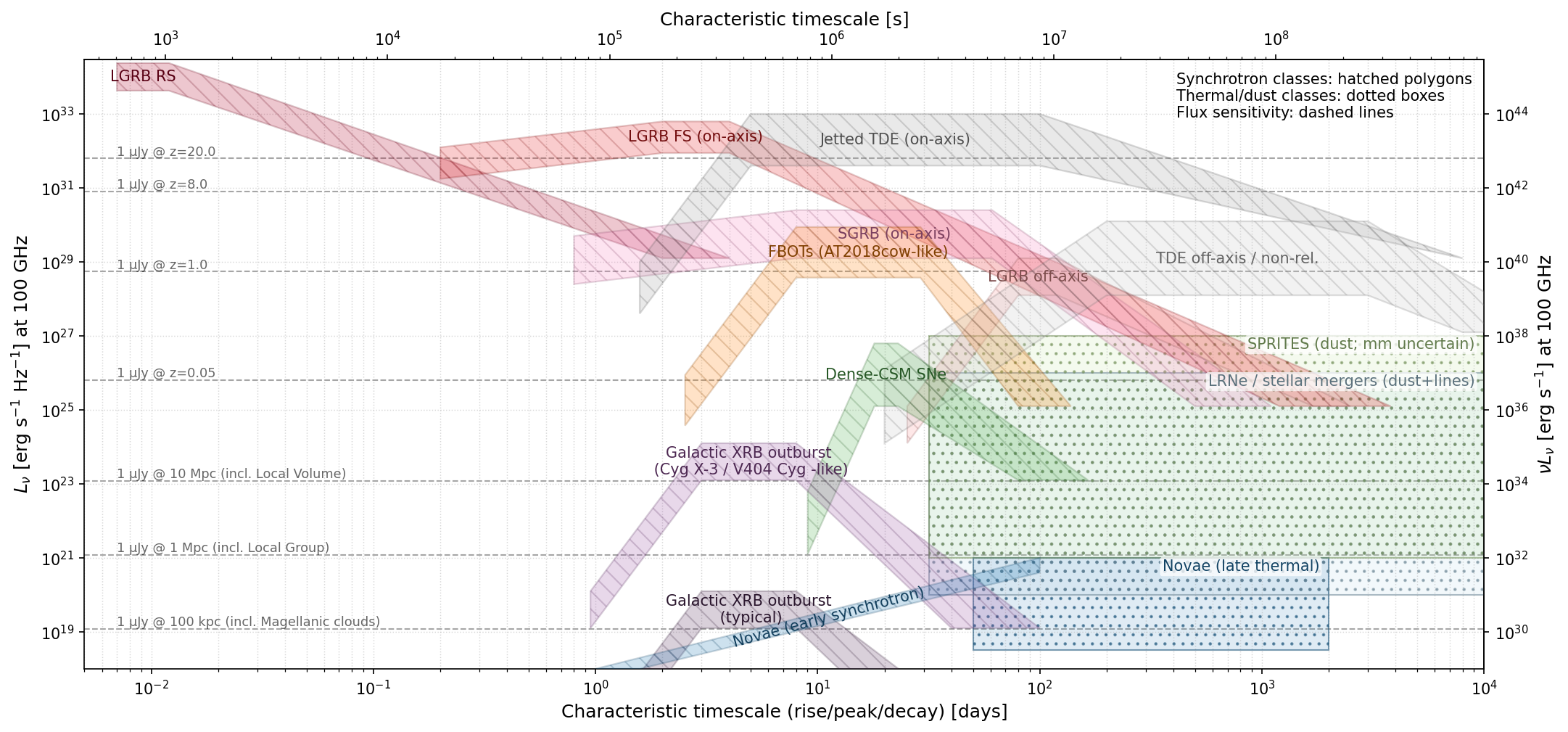}
    \caption{Millimeter luminosity (at 100 GHz) vs.\ event duration (rise/peak/decay) for astrophysical transients. A representative range is included for on-axis long-GRB reverse shock (LGRB RS), forward shock (LGRB FS), off-axis long-GRB, and on-axis short-GRB based on modeling; jetted on-axis TDEs, and FBOTs based on data \citep[][and references therein]{eftekhari22}; dense-CSM SNe based on data \citep{Maeda2021,Nayana2025}; Galactic XRBs based on data (\citet[][for outbursts]{Tetarenko2017,Koljonen2018} and \citet{Elford2026} for typical hard state XRBs); novae based on data \citep{chomiuk21}; and LRNe/SPRITES based on data and expectations \citep[][and references therein]{Kaminski2026}. Horizontal dashed lines show 1~$\mu$Jy flux sensitivity at selected distances (attainable with ALMA2040 in $\lesssim$1-hour exposure at 35--230 GHz).}
    \label{fig:zoo}
\end{figure}

Figure~\ref{fig:zoo} summarizes where transient classes sit in millimeter luminosity--timescale space, highlighting both the diversity of physical regimes accessible at mm wavelengths and the observational requirements implied by fast and faint transients.

\vspace{-0.5cm}

\section{Science Case for the 2040's}

By the 2040s, time-domain astrophysics will be transformed by sensitive mm surveys, advanced GW facilities and routine discovery of faint, fast-evolving extragalactic transients. ALMA2040 converts these discoveries into quantitative constraints on explosion physics. These include early multi-band snapshots that capture the evolving turnover when diagnostics are most constraining, high-cadence monitoring that tracks curvature, breaks and component evolution without variability-induced biases, and (where feasible) imaging/size constraints that reduce model degeneracies. Because mm emission often escapes when cm emission is absorbed and optical emission is reprocessed, mm follow-up is uniquely constraining for fast and obscured explosions. 

\vspace{0.1cm} \noindent \textbf{Multi-messenger connections (neutrinos and beyond)\,\,}
In the 2040s, next-generation neutrino facilities such as IceCube-Gen2 and KM3NeT are expected to deliver more frequent astrophysical neutrino alerts, often with localizations too coarse for direct interferometric tiling. Once credible counterpart candidates are identified by wide-field facilities, ALMA2040 provides physical inputs required by hadronic interaction and particle-acceleration models: outflow/shock energetics, circum-source density profiles and magnetic-field strength/ordering (via spectral evolution and polarimetry), as well as evidence for multiple emission components (e.g., reverse shocks or stratified ejecta). These measurements constrain the efficiency of $pp$ and $p\gamma$ interactions and the maximum particle energies, strengthening inference for candidate neutrino-producing transients including TDEs, interaction-powered SNe, FBOTs and GRBs.

Multi-messenger triggers (neutrino/GW/high-energy) will often have large error regions compared to a 12-m primary beam, requiring wide-field facilities and brokers to reduce the search area and identify candidates. Rapid ALMA2040 snapshots can then deliver sub-arcsecond positions and an initial SED. Figure~\ref{fig:zoo} illustrates why both sensitivity and rapid response matter: the fastest events require early-time snapshots and dense cadence, while the fainter sources require order-of-magnitude gains in sensitivity for routine physical characterization. 

\vspace{0.1cm} \noindent \textbf{Representative science drivers\,\,}
\underline{\textit{Dense-CSM SNe:}} 
Prompt, multi-band mm follow-up probes early turnovers and disentangles synchrotron self-absorption, free--free absorption and emerging dust emission during the critical first days to weeks, helping reconstruct the CSM density scale and radial structure \citep[e.g.,][]{chevalier98,weiler02,Maeda2021,Nayana2025}. With sufficient sensitivity at the highest frequencies and extended baselines, ALMA2040 could resolve a Galactic supernova on day timescales at the highest frequencies, enabling time-resolved imaging (``movies'') of shock expansion and dust formation.

\underline{\textit{GRBs/FBOTs/TDEs:}} 
Early GRB afterglows can evolve on minutes--hour timescales \citep{Perley2014,Laskar2013}), while jetted TDEs and FBOT likely exhibit stratified outflows evolving on days--weeks timescales \citep[e.g.,][]{ho19}. Rapid, simultaneous multi-band monitoring is required to capture evolving spectral peaks and to constrain electron distributions, energies and outflow structure \citep{margalit24}. These capabilities also enable interpretation of Laser Interferometer Space Antenna (LISA)-era transients, which will open discovery space for massive black hole mergers and related electromagnetic counterparts in gas-rich environments and will probe the demographics of lower-mass supermassive black holes relevant to TDE scenarios \citep[e.g.,][]{AmaroSeoane2007,Kulkarni2016,Bode2010,Milosavljevic2005,Metzger2016}.
  
\underline{\textit{TDE hosts:}}
Converting mm-inferred outflow parameters into absolute Eddington ratios and jet efficiencies requires emission-independent measurements of central black hole masses. ALMA dynamical masses from molecular-gas kinematics are now well established \citep[e.g.,][]{Davis2013,Barth2016,Boizelle2019,Nguyen2021} and are particularly valuable for low-mass and post-starburst TDE hosts \citep[e.g.,][]{Arcavi2014,French2016} where $M$--$\sigma$ extrapolations are uncertain \citep[e.g.,][]{Kormendy2013,Greene2020,Reines2015}.

\underline{\textit{XRBs:}}
An order-of-magnitude gain in sensitivity enables routine detections and spectral characterization of XRB jets in the Magellanic Clouds (well-constrained distances and metallicities) and extends the brightest events to the Local Volume. For microquasar outbursts, $S_\nu \sim 1$--10\,Jy at $\nu\sim$100\,GHz \citep{Tetarenko2017,Koljonen2018} corresponds to
$L_\nu \approx 1.2\times10^{23}\,(S_\nu/{\rm Jy})\,(d/10\,{\rm kpc})^2\ {\rm erg\,s^{-1}\,Hz^{-1}}$, and sub-arcsecond localization will connect these events to their host environments and stellar populations. 

\underline{\textit{Novae:}}
Improved angular resolution and sensitivity are required to separate free--free, synchrotron and dust components in the earliest days of an eruption, yielding accurate determinations of ejecta masses, velocities and shock energetics \citep[e.g.,][]{chomiuk21,Vlasov2016,Linford2017}. Novae in the Magellanic Clouds provide a controlled test of how composition regulates thermonuclear runaway, mass ejection and shock interaction. In this regime, recurrent nova eruptions offer a lower-energy analogue of supernova shock evolution, with ejecta--circumbinary interaction unfolding on dramatically shorter timescales \citep[e.g.,][]{Starrfield2016,Darnley2012}.

\underline{\textit{Stellar mergers:}}
ALMA2040 can characterize the morphologies and molecular compositions of merger remnants across the Galaxy and the Local Group, offering insight into mass-loss channels during stellar coalescence and quantifying their role in dust production. Time-resolved continuum and spectral-line observations of LRNe and other dust-embedded transients will reveal the formation of structures such as disks, tori, jets and wide-angle outflows in real time, providing direct constraints on angular-momentum redistribution and on the physical conditions enabling efficient molecule and dust formation \citep[e.g.,][]{KamiPN,Kaminski2026}. 

\underline{\textit{GW counterparts:}}
Early mm follow-up (100--300~GHz) captures phases less affected by self-absorption and tracks rapidly evolving peaks with multi-band measurements \citep[e.g.,][]{Granot2002,Hallinan2017,Mooley2018,Margutti2018}. These data strongly constrain ejecta/outflow structure, jet angular structure and circum-merger environments, and can help constrain the nature of the compact remnant, particularly through signatures of late-time energy injection or non-thermal emission from a long-lived central engine \citep[e.g.,][]{Lazzati2018,Troja2018,Mukhopadhyay2025}.

\section{ALMA2040: Technical Requirements}

%Summarize the technical requirements here (e.g., angular resolution, spectral resolution, band, bandwidth, sensitivity).
%\subsection{Simulations with GHOST}
%Describe the configurations used to obtain the mock observations from GHOST at \url{https://alma-ghost.jb.man.ac.uk} and present the results.
\begin{table*}
\centering
\caption{ALMA2040 capabilities mapped to key transient-science deliverables and representative source classes. Requirements summarize approximate gains of ALMA-WSU and ALMA2040 relative to current ALMA.}
\label{tab:req_science}
\begin{tabular}{p{2.2cm} p{6.8cm} p{4.2cm} >{\centering\arraybackslash}p{1.7cm}}
\hline
\textbf{Capability} & \textbf{Science questions / deliverables} & \textbf{Classes impacted} & \textbf{Req.}\\
\hline
Order-of-magnitude continuum sensitivity gain &
Routine early-time detections for faint/fast extragalactic events; track spectral peak/turnover evolution to infer energy and density profiles; expand samples beyond rare nearby cases &
Dense-CSM SNe, off-axis GRBs, FBOTs, non-jetted TDEs, GW counterparts, extragalactic XRB flares &
\statS{2}{10-20} \\\hline
\vspace{-0.1cm}
Wide instantaneous bandwidth &
\vspace{-0.1cm}
Snapshot SEDs without variability-induced spectral biases; detect/quantify spectral curvature and breaks testing shock acceleration/cooling; separate synchrotron/free--free/dust &
\vspace{-0.1cm}
GRBs (early), FBOTs, jetted/non-jetted TDEs, interaction-powered SNe, novae &
\vspace{-0.1cm}
\statBW{4}{64}\\\hline
\vspace{-0.1cm}
Rapid-response &
\vspace{-0.1cm}
Capture rapidly evolving early phases (reverse shocks; evolving turnovers); near-simultaneous multi-band ToOs &
\vspace{-0.1cm}
GRBs, FBOTs, GW afterglows, early interaction-powered SNe, neutrino-alert counterparts &
\vspace{-0.1cm}
\statQ{n/a}{mins--hr}\\\hline
\vspace{-0.1cm}
Sub-arrays &
\vspace{-0.1cm}
Efficient monitoring of bright or multiple targets discovered by surveys/brokers; dedicated VLBI array readiness &
\vspace{-0.1cm}
All fast transients; high discovery-rate regimes; VLBI targets &
\vspace{-0.1cm}
\statQ{limited}{routine}\\\hline
\vspace{-0.1cm}
Polarimetry &
\vspace{-0.1cm}
Magnetic-field ordering, jet magnetization, geometry constraints; distinguish ordered jet cores vs.\ turbulence-dominated shocks; test microphysical models &
\vspace{-0.1cm}
GRBs, jetted TDEs, FBOTs, XRB jets, some SNe &
\vspace{-0.1cm}
\statQ{yes}{ToO-capable}\\\hline
\vspace{-0.1cm}
Order-of-magnitude spectral-line sensitivity &
\vspace{-0.1cm}
Gas kinematics/chemistry around transients; CSM/circumnuclear structure; molecule/dust formation and shocked gas in obscured systems &
\vspace{-0.1cm}
Interaction-powered SNe (CSM), LRNe/mergers, novae, TDE environments &
\vspace{-0.1cm}
\statS{2}{10}\\\hline
\vspace{-0.1cm}
Extended baselines &
\vspace{-0.1cm}
Direct size/expansion constraints to break energy--density--microphysics degeneracies; resolve/remnant morphology and asymmetries; time-resolved imaging where feasible &
\vspace{-0.1cm}
Galactic (SNe, novae, mergers); nearest extragalactic targets &
\vspace{-0.1cm}
\statQ{n/a}{3$\times$}\\\hline
\vspace{-0.1cm}
mm-VLBI: full-array phasing + multi-band + rapid ToO &
\vspace{-0.1cm}
Resolve jet structure directly in selected bright extragalactic transients; unique geometric constraints beyond light-curve modeling &
\vspace{-0.1cm}
GRBs, TDEs, (some) SNe/FBOTs &
\vspace{-0.1cm}
\statQ{limited}{routine}\\
\hline
\end{tabular}
\end{table*}

Table~\ref{tab:req_science} maps key transient-science questions to the ALMA2040 capabilities required to answer them and summarizes approximate gains relative to current ALMA. Below we highlight the points most relevant for time-domain operation. 

\vspace{0.1cm} \noindent \textbf{Response time requirements across transient classes\,\,}
Rapid-response modes that place the array on source within minutes to hours of an external trigger, combined with flexible dynamic scheduling and sub-arraying, are required to capture the fast evolution while managing the high discovery rates expected from surveys. The response time requirement depends on source class (Figure~\ref{fig:zoo}): (i) \textit{minutes--hours} for early GRB afterglows (including reverse-shock signatures) and the fastest engine-powered transients; (ii) \textit{hours--day} for capturing early spectral turnovers in dense-CSM SNe and rapidly evolving FBOTs; and (iii) \textit{days} for many TDEs, LRNe/SPRITES and slower-evolving obscured transients. Across all classes, near-simultaneous multi-band capability is required to avoid aliasing intrinsic variability into apparent spectral curvature.

\vspace{0.1cm} \noindent \textbf{Imaging, baselines and mm-VLBI\,\,} Extended baselines within ALMA2040 will primarily enable resolved imaging and expansion measurements for Galactic events (novae, stellar mergers and a Galactic supernova) and, in favorable cases, the nearest Local Group transients; for most extragalactic explosions, the emission region will remain unresolved, but meaningful size and brightness-temperature limits can still be obtained. In parallel, ALMA2040 will play a transformative role as a phased element in global mm-VLBI arrays (e.g., the Event Horizon Telescope, EHT; see also the companion ALMA2040 white paper \textit{The transformative power of ALMA2040 for mm-VLBI}), enabling $\mu$as-scale imaging of relativistic outflows in selected bright extragalactic transients (e.g., GRBs, TDEs and some SNe/FBOTs) and providing direct geometric constraints inaccessible from light curves alone.

\vspace{0.1cm} \noindent \textbf{Polarimetry and calibration\,\,}
Polarization observations can distinguish between highly ordered magnetic fields, turbulence-dominated shocks and magnetized jet cores, and can provide an independent constraint on geometry when combined with spectral evolution. To enable such diagnostics for fast transients, polarization observations must be available in rapid-response modes with sufficient sensitivity to measure or constrain polarization fractions at the percent level for typical ToO flux densities, and with stable calibration suitable for time-variable sources. Likewise, because many transients evolve on minute-to-hour timescales, multi-band spectral sampling through wide instantaneous bandwidth is required.

\vspace{0.1cm} \noindent \textbf{Time-domain infrastructure\,\,}
Because the most constraining phases of many mm transients occur early and evolve rapidly, ALMA2040 must function as a true time-domain facility with automated trigger ingestion from multi-messenger and survey brokers, rapid feasibility assessment (weather, configuration, frequency), low-latency calibration and imaging for transient verification, and scheduling modes that support monitoring across multiple targets and bands. Such infrastructure maximizes scientific return per ToO by enabling early data with high quality (positions and broad-band SEDs) and by allowing the observing strategy to adapt in real time as spectral peaks move and components emerge or fade.

%
%\section{Synergies with current and future facilities}
%This section is optional. 
\section{Synergies with current and future facilities}

Looking ahead, high-cadence mm surveys conducted by CMB survey experiments in the 2030s, such as the Simons Observatory \citep{abitpol25}, and their successors in the 2040s, will uncover a wealth of short-lived explosive transients and possibly entirely new phenomena \citep{eftekhari22,metzger15}. The coarse localization of such surveys will require a rapid, high-resolution follow-up likely aided by wide-field sub-mm/mm facilities (e.g., AtLAST) that can efficiently identify and/or prioritize candidate counterparts for targeted ALMA2040 observations. 

Moreover, by the 2040s the transient landscape will be increasingly multi-messenger: next-generation GW detectors (such as Einstein Telescope and Cosmic Explorer) and neutrino facilities (such as IceCube-Gen2 and KM3NeT) are expected to deliver triggers for compact object mergers and relativistic outflows on a routine basis, and the mm band will provide a unique and physically direct window into the associated shocks, jets and obscured environments.

Wide-field optical and near-infrared surveys in the Rubin/LSST era (and next-generation time-domain facilities operating in the 2040s, e.g., Argus Array) will provide discovery, classification and host-galaxy context for many transients, and will be essential for reducing large multi-messenger error regions to a manageable set of candidate counterparts for targeted mm follow-up. High-energy facilities, including wide-field X-ray monitors and the Cherenkov Telescope Array Observatory (CTAO), will provide prompt timing and energetics constraints for relativistic explosions, enabling rapid ALMA2040 response during the phases when mm spectra evolve most rapidly. At longer wavelengths, next-generation cm facilities (e.g., SKAO and/or ngVLA) will complement ALMA2040 by tracing late-time, optically thin synchrotron emission and low-frequency absorption, improving calorimetry and constraining ambient densities over a wider range of radii. Together, these facilities provide the triggers, classifications and multi-band baselines that allow ALMA2040 to focus on the mm/sub-mm measurements that most directly probe shock physics, geometry and obscured environments.

In particular, the extremely large telescopes (such as ELT, GMT, TMT) will provide capabilities that complement the mm interferometry: dynamical mass measurements and mid-infrared access to dust-enshrouded ejecta. Once a transient has faded, integral-field spectroscopy can deliver nuclear stellar kinematics for nearby TDE hosts, enabling emission-independent black hole mass estimates that anchor mm-derived outflow energetics in absolute Eddington units. ELT observations can also resolve the local explosion environment (stellar population, metallicity and ionization conditions), providing the context into which the shocks traced by ALMA2040 expand. For dust-rich transients such as LRNe and stellar mergers, ELT mid-IR instrumentation will trace warm-dust emission and key molecular features (e.g., CO, SiO), complementing ALMA2040 maps of remnant morphology and cold gas with constraints on dust chemistry and temperature. Together, ALMA2040 and ``ELTs'' provide a coordinated view of shocks, dust and host-galaxy context that is difficult to achieve with any single facility.
 
\section*{Acknowledgements}
We sincerely thank Tarraneh Eftekhari (Northwestern University) and Mainak Mukhopadhyay (University of Chicago) for reviewing this paper. This work was developed as part of the broader ALMA2040 community effort. We acknowledge the Lorentz Center and NOVA for supporting the community workshops and coordination activities that helped initiate and shape the ALMA2040 White Paper process.\\

\bibliographystyle{apj_mod}
\noindent {\bibliography{references}}
\end{document}